\documentclass[sigconf]{acmart}
\AtBeginDocument{%
  }

\usepackage{enumitem}
\usepackage{graphicx}
\usepackage{adjustbox}
\usepackage{pifont}
\usepackage{array}
\usepackage{xcolor}
\usepackage{multirow}
\usepackage[table]{xcolor}
\usepackage{balance}
\usepackage{tcolorbox}
\usepackage{listings}
\tcbuselibrary{listingsutf8}
\usepackage{listings}
\definecolor{lightgray}{gray}{0.9}
\newcommand{\filledC}{\ding{108}} %
\newcommand{\emptyC}{\ding{109}} 

\copyrightyear{2025}
\acmYear{2025}
\setcopyright{cc}
\setcctype{by}
\acmConference[CCS '25]{Proceedings of the 2025 ACM SIGSAC Conference on Computer and Communications Security}{October 13--17, 2025}{Taipei, Taiwan}
\acmBooktitle{Proceedings of the 2025 ACM SIGSAC Conference on Computer and Communications Security (CCS '25), October 13--17, 2025, Taipei, Taiwan}\acmDOI{10.1145/3719027.3765072}
\acmISBN{979-8-4007-1525-9/2025/10}

\begin{document}

\title[Layered, Overlapping, and Inconsistent Privacy Policies and Controls]{Layered, Overlapping, and Inconsistent: A Large-Scale Analysis of the Multiple Privacy Policies and Controls of U.S. Banks}

\author{Lu Xian}
\orcid{0000-0001-8120-1012}
\affiliation{%
  \institution{University of Michigan}
  \city{Ann Arbor}
  \country{USA}}
\email{xianl@umich.edu}

\author{Van Hong Tran}
\orcid{0000-0002-0835-8598}
\affiliation{%
  \institution{University of Chicago}
  \city{Chicago}
  \country{USA}}
  \email{tranv@uchicago.edu}

\author{Lauren Lee}
\orcid{0009-0002-2287-9127}
\affiliation{%
  \institution{University of Michigan}
  \city{Ann Arbor}
  \country{USA}}
\email{laurnlee@umich.edu}

\author{Meera Kumar}
\orcid{0009-0007-3433-4067}
\affiliation{%
  \institution{University of Michigan}
  \city{Ann Arbor}
  \country{USA}}
\email{meerakmr@umich.edu}

\author{Yichen Zhang}
\orcid{0009-0004-9078-794X}
\affiliation{%
  \institution{University of Wisconsin-Madison}
  \city{Madison}
  \country{USA}}
\email{zhang2276@wisc.edu}

\author{Florian Schaub}
\orcid{0000-0003-1039-7155}
\affiliation{%
  \institution{University of Michigan}
  \city{Ann Arbor}
  \country{USA}}
\email{fschaub@umich.edu}

\renewcommand{\shortauthors}{Lu Xian et al.}

\begin{abstract}
Privacy policies are often complex. An exception is the two-page standardized notice that U.S. financial institutions must provide under the Gramm-Leach-Bliley Act (GLBA). However, banks now operate websites, mobile apps, and other services that involve complex data sharing practices that require additional privacy notices and do-not-sell opt-outs. We conducted a large-scale analysis of how U.S. banks implement privacy policies and controls in response to GLBA; other federal privacy policy requirements; and the California Consumer Privacy Act (CCPA), a key example for U.S. state privacy laws. We focused on the disclosure and control of a set of especially privacy-invasive practices: third-party data sharing for marketing-related purposes. We collected privacy policies for the 2,067 largest U.S. banks, 45.2\% of which provided multiple policies. Across disclosures and controls for the \textit{same} bank, we identified frequent, concerning inconsistencies---53.8\% of banks with multiple privacy policies indicated in GLBA notices that they do not share with third parties but disclosed sharing in other policies. This multiplicity of policies, with the inconsistencies it causes, may create consumer confusion and undermine the transparency goals of the very laws that require them. Our findings call into question whether current policy requirements, such as the GLBA notice, are achieving their intended goals in today's online banking landscape. We discuss potential avenues for reforming and harmonizing privacy policies and control requirements across federal and state laws.
\end{abstract}



\received{14 April 2025}
\received[accepted]{2 July 2025}

\maketitle

\section{Introduction}

In the U.S., consumer privacy laws rely on the “notice and choice” framework \cite{FTC2008}, which requires businesses to provide privacy notices and opt-out choices, so that consumers can understand data practices and exercise privacy choices. Numerous studies have shown that this framework neither effectively informs consumers nor supports their ability to express choices \cite[e.g.,][]{Cranor2012,cranor2024notice, cate2016failure}. 
Design efforts have sought to make privacy information more accessible and comprehensible~\cite{schaub2015design, kelley2009nutrition, zhang2022usable}.
A notable example is the standardized short-form notice (see Figure~\ref{fig:glba-model}) that U.S. financial institutions must provide under the federal Gramm-Leach-Bliley Act (GLBA)~\cite{FTC2002,garrison2012designing}.

However, the GLBA narrowly applies to ``nonpublic personal information'' relating to financial products or services \cite{FTC2002, uscode15_6801}.
Today, financial institutions often collect and process additional data types through mobile apps and online services, which may require further privacy notices and controls, such as a general privacy policy.  
In addition, U.S. state privacy laws, such as the California Consumer Privacy Act (CCPA),
impose further privacy disclosure and opt-out requirements on institutions that do business in the respective state. 

Given the range of privacy regulations and the differences in their scope, definitions, and requirements, U.S. banks have started providing multiple notices and opt-out choices in addition to the GLBA short-form notice. 
We conducted a large-scale analysis of the privacy policies and respective privacy controls for the 2,073 largest commercial banks in the U.S., which collectively hold 97.3\% of all assets of FDIC-insured commercial banks.  
We examined whether a given bank provides multiple privacy policies (GLBA, general, mobile, cookie, and CCPA policies) and controls (GLBA, cookie, and CCPA opt-outs), and whether inconsistencies exist across these policies that could mislead or confuse consumers. 
We focused specifically on \textit{third-party data sharing for marketing}, which we considered to include both marketing/advertising and analytics/research purposes more broadly, along with the related opt-outs, because people often find these practices concerning~\cite{rao_expecting_2016, weinshel_oh_2019, al2020most}, as they constitute violations of contextual integrity~\cite{nissenbaum_contextual_2011}. 
Our \textbf{research questions} were:
\noindent \begin{itemize}
\item[RQ1] How many privacy policies are consumers likely to encounter for a given U.S. bank? What do their length and readability reveal about the effort required for consumers to understand a bank’s data practices? 
\item[RQ2] What do a bank's multiple privacy policies, provided in response to different regulations, 
disclose about third-party sharing practices regarding marketing and advertising purposes? Are these disclosures consistent across multiple policies provided by the \textit{same} bank?  
\item[RQ3] How do banks provide privacy opt-outs regarding third-party sharing for marketing purposes, as required by different regulations? 
\end{itemize}

\textbf{Summary of findings.} 
We found privacy policies for 2,069 banks, 44.2\% of which provided multiple privacy policies, most commonly a GLBA notice in combination with a general or mobile privacy policy. 
The combined privacy policy content for each bank requires a median reading level equivalent to a college education, with larger banks providing lengthier and more difficult-to-read text. 
We identified two distinct types of inconsistency across a bank's privacy disclosures. Concerningly, many banks appear more privacy protective in the GLBA notice than in other disclosures: 53.8\% of banks with multiple privacy policies indicated in their GLBA notice that they would not share data with third parties for marketing purposes, yet disclosed such sharing in their other policies. 
The opposite inconsistency type was less concerning, though still notable: 36.8\% of banks with multiple privacy policies indicated sharing in the GLBA notice, while in CCPA-related disclosures they stated that they do not sell/share data, possibly reflecting CCPA's stricter limits on data sharing.
Many banks that disclosed data-sharing practices did not offer the required opt-outs, and most banks had third-party marketing cookies on their websites without disclosing them or offering opt-out controls. 

The multiplicity of policies and identified inconsistencies shows that for many banks the short-form GLBA notice no longer provides a full representation of their third-party sharing practices. Consumers must now navigate multiple privacy documents, with the documents' varying scopes in mind, to learn about and manage their data and understand their privacy choices. 
Our findings highlight that the narrowly-scoped GLBA notice may mislead consumers, and that the layering of different disclosure requirements can undermine the transparency goals of the very laws that require them. 
We discuss concrete opportunities for regulatory reform that could reduce duplication, resolve inconsistencies across notices, and ultimately make privacy information more accessible and actionable for consumers.

\section{Background}\label{sec:background}

We discuss privacy notice and opt-out requirements of GLBA, other privacy notice requirements, and California privacy laws.

\subsection{The Gramm-Leach-Bliley Act}

\begin{figure}[t]
    \centering
    \includegraphics[width=\columnwidth]{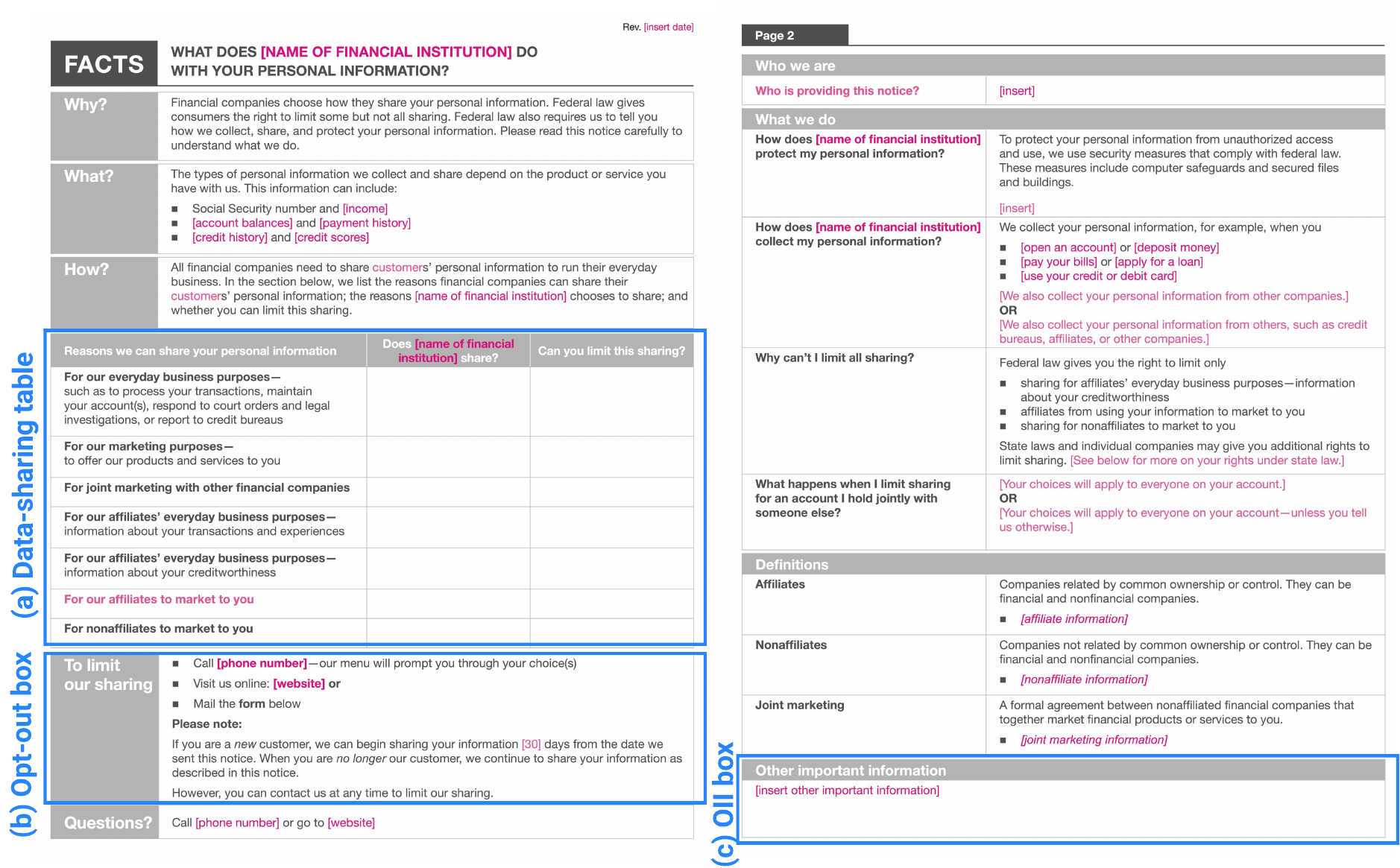} 
    \caption{2-page GLBA model privacy form \cite{SEC34-61003}. We analyzed the (a) data-sharing table, (b) opt-out box, (c) OII box (blue annotations added for clarity, not part of the template).}
    \label{fig:glba-model}
\end{figure}

The Gramm-Leach-Bliley Act of 1999 (GLBA) requires financial institutions to disclose their information collection and sharing practices annually 
as well as inform customers of their right to opt out of certain sharing practices \cite{PL106-102}. 
The GLBA narrowly covers ``nonpublic personal information'' 
related to providing financial products or services~\cite{FTC2002, uscode15_6801}, such as a consumer’s name, income, and social security number.
A two-page GLBA model privacy form
~\cite{SEC34-61003} prescribes a standardized layout with pink-bracketed text to be customized for an institution’s data practices (see Figure~\ref{fig:glba-model}).
Although GLBA notices were historically delivered by postal mail, 
many banks now send these notices via email and provide them on their websites in PDF format. 

The model privacy form highlights how bank customers' financial information is collected, shared, and protected in a table-based format. It was consumer tested for judgment quality, perceptual accuracy, and reading ease \cite{kleimann2006evolution, garrison2012designing, LevyHastak2008}.
On the first page, the ``Why?,'' ``What?,'' and ``How?'' text boxes summarize disclosure requirements as well as five examples of personal information types, chosen from a pre-defined list \cite{CFPB_Regulation_P_Appendix}, that the institution collects and shares. 
Our analysis focuses on the data-sharing table (\textit{data-sharing table}; see Figure~\ref{fig:glba-model}(a)), which lists at most seven pre-defined data-sharing practices distinguished by sharing purposes and informs consumers about whether they can limit each sharing. Among the seven purposes, the ``For our affiliates to market to you'' purpose may be omitted in certain cases, while the others are required to be listed.

An institution must provide an opt-out option for the last three of these purposes.  
Respective opt-out instructions are described in a ``To limit our sharing'' box on the first page (\textit{opt-out box}, see Figure~\ref{fig:glba-model}(b)).
We found that many financial institutions use the ``Other important information'' box (\textit{OII box}, see Figure~\ref{fig:glba-model}(c)) on the second page to point to additional data practice disclosures and opt-out rights under state-level privacy laws. 
The disclosures refer to affiliates, which are financial or nonfinancial companies under common ownership or control with the disclosing institution; nonaffiliates, which are unrelated third parties; and joint marketers, which can be affiliates and nonaffiliates~\cite{CFPB_Regulation_P_Appendix}. 

\subsection{Other Privacy Notice Requirements}

As part of its mandate to protect consumers from businesses engaging in unfair or deceptive commercial practices, the Federal Trade Commission (FTC) has provided guidance recommending privacy policies for websites and mobile apps~\cite{FTC_Mobile_App_Guide, FTC2012Privacy} to ensure data practices are transparent.  
State-level regulations, such as the CCPA \cite{CalBPC_Ch22}, also require commercial websites and online services, including banks, to post privacy policies detailing their data collection and sharing practices. 
Banks may also voluntarily provide cookie or privacy notices to align with international regulations such as the EU's General Data Protection Regulation or ePrivacy Directive. Compared to GLBA, these additional notices generally have fewer specific content requirements and thus vary widely in content, structure, and format.

\subsection{California Privacy Laws}

California has a state privacy law for the financial industry, the California Financial Information Privacy Act (CalFIPA)~\cite{california_financial_code_4050}, which requires explicit consent for certain types of marketing-related third-party sharing. 
In addition, the California Consumer Privacy Act (CCPA) is a comprehensive privacy law encompassing almost all industries.
Businesses are subject to CCPA if they operate in California; collect, sell, or share consumer personal information; and meet certain thresholds \cite{ccpa_regulations_2023,statute}. 
CCPA defines ``personal information'' more broadly than GLBA's ``nonpublic personal information,'' covering any data that identifies, relates to, or can reasonably be linked to a consumer or their household.
This includes consumers' names, social security numbers, email addresses, records of purchased products, browsing history, location data, etc.

Under CCPA, businesses that sell or share personal information with third parties are required to provide a notice of the right to 
opt-out of the sale/sharing and a method for doing so~\cite{california_civil_code_1798_140}. 
Sharing under CCPA specifically refers to disclosures made for ``cross-context behavioral advertising'' with third parties, excluding service providers and contractors
~\cite{california_civil_code_1798_140}. 
Compared to GLBA, 
CCPA specifies required content in detail but does not mandate notice formats. 
We observed that some banks included the notice of the right to opt out of sale/sharing within their general privacy policy, while others used a standalone CCPA policy, both with varying formats. 

Businesses must also respect opt-out preference signals~\cite{ccpa_reg}, like the Global Privacy Control (GPC) signal \cite{gpc}, 
and provide at least one method for consumers to opt out of the sale or sharing of their personal information. This can be done either through a ``Do Not Sell or Share My Personal Information'' link or an alternative opt-out link (``Your Privacy Choices''), which we refer to collectively as CCPA opt-out links. Alternatively, if businesses honor opt-out preference signals in a frictionless manner, meaning the request is automatically processed and consumers are opted out of all sale/sharing of personal information without any further action \cite{ccpa_reg}, they are exempt from providing an opt-out link. \looseness=-1

The CCPA includes a carve-out for data covered under GLBA, meaning that personal information collected, processed, sold, or disclosed pursuant to the GLBA (i.e., as part of financial transactions) is exempt from CCPA requirements. 
However, banks are subject to CCPA for other personal information they process. 
For example, website tracking data, such as cookies that monitor browsing behavior for retargeting purposes, is outside of GLBA's scope but covered under CCPA. Similarly, a bank website's use of third-party tracking scripts for cross-context behavioral advertising falls under CCPA. Partial overlaps and gaps like these create a patchwork in which different information of the same consumer held by a bank is governed by multiple regulations, with associated data practices disclosed in different privacy policies.


\section{Related Work}

We discuss related work on inconsistencies in privacy disclosures and practices, readability of privacy policies, usability of privacy controls, and legal compliance analysis.

\textbf{Inconsistencies in privacy disclosures and practices.} 
Prior research has uncovered significant inconsistencies and contradictions in businesses' privacy disclosures and practices, both within privacy policies regarding stated data collection and sharing practices, and between privacy policies and actual data handling. For example, Andow et al. \cite{andow2019policylint} analyzed the privacy policies of 11,430 apps and found that 14.2\% contained contradictions, potentially indicating misleading statements. The study highlighted several concerning patterns, such as the use of misleading language, attempts to redefine commonly understood terms, and the concealment of tracking data through data sharing or collection methods that could indirectly reveal sensitive information. Studies have also revealed concerning inconsistencies between mobile apps’ stated privacy policies and their actual data practices.
Bui et al. \cite{bui2021consistency} and Slavin et al. \cite{slavin2016toward} found that nearly 70\% of apps failed to align their data practices with their stated privacy policies and almost always over-collected consumer personal data. 
Similarly, Nguyen et al. \cite{nguyen2022freely} showed that some apps continued transmitting user data even after users explicitly opted out, directly violating consumer expectations. Andow et al. \cite{andow_actions_2020} characterized such ``flow-to-policy'' inconsistencies in terms of data type and recipient and found many of them in mobile apps.
These findings underscore a recurring pattern of misleading privacy disclosures and non-compliant data practices.
Building on prior work that primarily examined specific and separate notices, our study analyzes inconsistencies among multiple privacy policies and opt-out choices provided by the same institution.

\textbf{Readability of privacy policies.}
Extensive research has examined the readability of privacy policies, identifying persistent clarity issues despite regulatory efforts, and finding that regulation has had a mixed effect on privacy policy transparency \cite{birrell2023sok}.
For example, Chen et al.'s \cite{chen2021fighting} analysis of the privacy policies of 95 popular websites found that despite the CCPA's mandate for clear privacy disclosures, privacy policies varied significantly in both the level of detail provided and in how key CCPA definitions like ``sale,'' ``valuable consideration,'' or ``business purpose'' are interpreted by businesses.  
Their survey found that many consumers found it difficult to fully understand how their data is collected and shared.
Similar concerns arise in studies examining the impact of the GDPR on privacy disclosures. Kretschmer et al. \cite{kretschmer2021cookie} and Degeling et al. \cite{degeling2018we} found that while transparency has improved since the introduction of the GDPR, usability challenges remain, especially in complex interface designs that limit user agency.
Wagner et al. \cite{wagner2023privacy} also found that privacy policies remain difficult to read, with legal reforms prompting only incremental improvements in readability. 
These findings suggest that despite regulatory intentions for transparency, privacy policies often remain too complex, inconsistent, or difficult to navigate, which limits their effectiveness in truly informing consumers~\cite{cranor2012necessary}. 
Rather than examining the full scope of privacy policies, we focus on consumers’ information needs related to exercising privacy controls---which typically pertain to limiting data sharing---and analyze how data-sharing practices are disclosed across privacy policies of the same bank.

\textbf{Usability of privacy controls.}
Usability issues of privacy controls meant to enable consumers to exercise their privacy rights can significantly impact consumer understanding and willingness to engage with them \cite{habib2020s, siebel2022impact, habib2021toggles}. 
Studies have found that clear, standardized, and prominently visible banners improve opt-out rates and enhance user satisfaction \cite{siebel2022impact}. In contrast, dark patterns---manipulative design tactics---can lead users, particularly those with lower digital literacy, to select less privacy-protective settings \cite{luguri2021shining, utz2019informed}. Despite their importance, privacy controls often present significant usability barriers \cite{utz2019informed}. Users frequently struggle to locate, understand, and effectively use these controls due to inconsistent placement, complex navigation, and a lack of clear instructions \cite{habib2019empirical}. Dark patterns also frequently appear in consent pop-ups \cite{nouwens, utz2019informed} and CCPA opt-out processes \cite{tran2024dark}, making privacy choices less accessible, more confusing, and unnecessarily burdensome. These findings suggest that poor usability and manipulative design often prevent consumers from effectively controlling their personal data. Our study found that most banks offer burdensome opt-out methods under the GLBA and that cookie controls vary widely in design and labeling, making it difficult for consumers to understand their purpose and exercise meaningful choice.

\textbf{Compliance with privacy regulations.}
Businesses' privacy policies and data practices often fail to meet regulatory requirements \cite{van2022setting, aziz2024johnny, tran2024measuring, tran2024dark, article}. 
Cranor et al.'s \cite{Cranor2016} large-scale analysis of financial institutions' GLBA notices found that, in 2014, many failed to provide required opt-out mechanisms, and some incorrectly stated that consumers could not limit certain types of data sharing that the GLBA permits them to restrict. Our analysis of GLBA notices found lower percentages of non-compliance than Cranor et al. reported a decade ago~\cite{Cranor2016}. We further expand on Cranor et al.'s focus on GLBA notices by investigating how GLBA notices relate to other, more recently prevalent privacy notices (e.g., those required by CCPA). To our knowledge, this is the first study to comparatively examine the concise, two-page GLBA notice with the broader set of privacy policies provided by the same banks. 

Studies examining CCPA and GDPR compliance \cite{tran2024measuring, o2021clear, article} reveal that businesses frequently fail to implement legally required opt-out links and cookie controls. Even when these mechanisms are provided, manipulative design tactics are often used to discourage consumers from exercising their rights, in direct violation of CCPA's restrictions \cite{tran2024dark}. Additionally, Aziz et al. \cite{aziz2024johnny} and Zimmeck et al. \cite{zimmeck2023usability} showed that automated privacy signals like Global Privacy Control (GPC), which are intended to provide consumers with an automated way to opt out, are frequently ignored by websites, despite legal mandates to honor them. Similarly, we found that some banks in our set ignored GPC signals even though they disclosed sale/sharing under the CCPA definition.
Beyond interface-level manipulations, compliance gaps persist in actual data practices. Studies by Matte et al. \cite{matte2020cookie} and Zhou et al. \cite{zhou2023policycomp} found that many apps and websites continue to collect personal data without proper disclosure, often using pre-selected consent options that subtly steer users toward agreement. 
The growing complexity of sector-specific and state-level privacy laws makes it increasingly difficult to maintain compliance and coherence. 
Our study examines the interplay between different compliance requirements through a consistency analysis, as consistency, clarity, and accessibility both within and across privacy documents provided by a single organization shape a consumer's understanding and control of the use of their data.


\section{Methods}

We collected the different privacy policies and third-party sharing opt-outs of the 2,073 largest U.S. banks, analyzed their statements regarding third-party sharing, and identified inconsistencies among a bank's different policies, opt-outs, and cookie practices.

\subsection{Data Collection}

\begin{figure*}[t]
    \centering
    \includegraphics[width=0.85\textwidth]{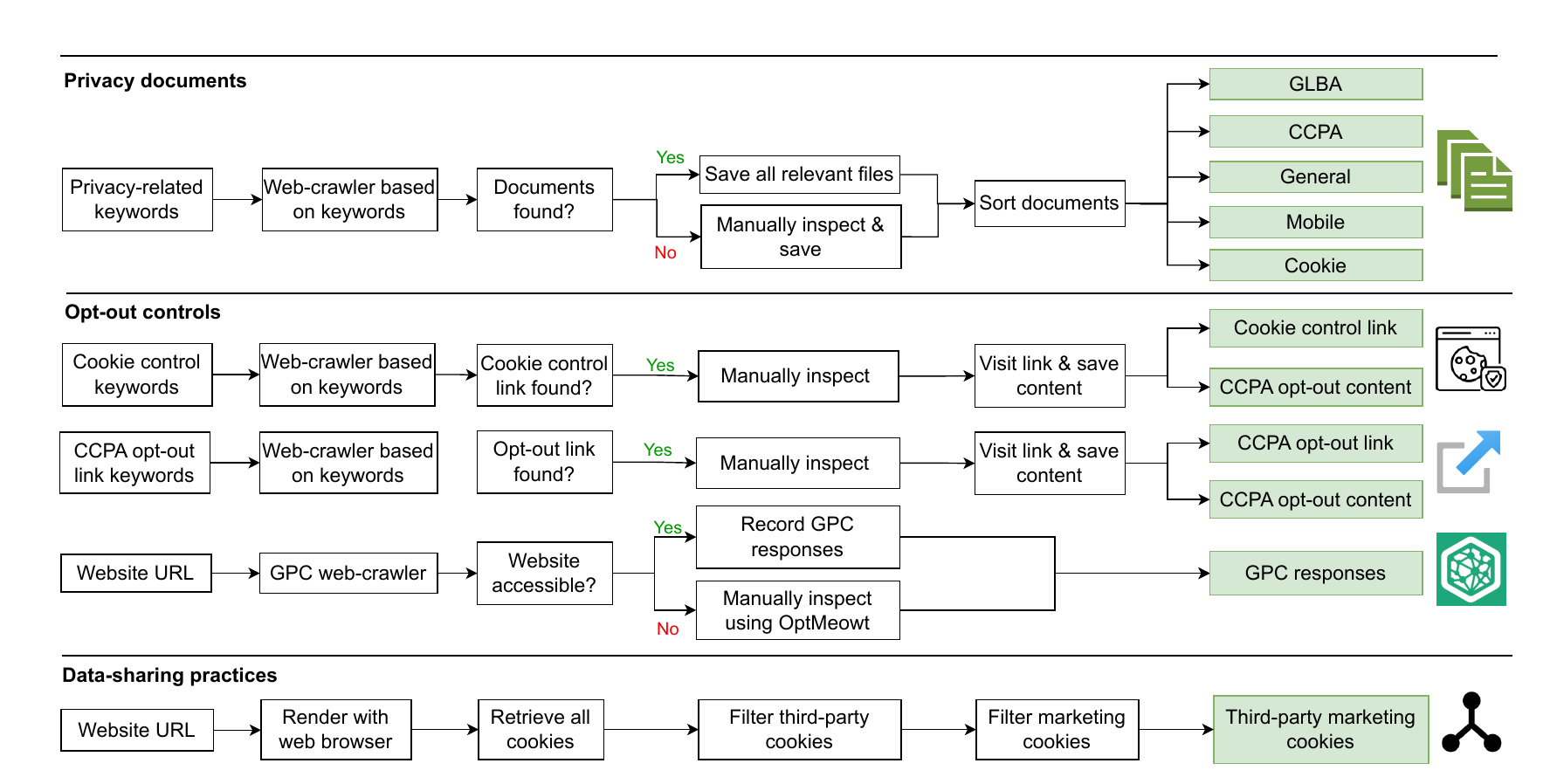} 
    \caption{Data collection pipeline.}
    \label{fig:data-collection-method}
\end{figure*}

We used a list of the largest commercial banks by the Federal Reserve \cite{federalreserve2024largebanks}, with consolidated assets exceeding \$300 million, each uniquely identified by its RSSD ID, a Federal Reserve-assigned identifier ($n$=2,129).\footnote{Consumer-oriented institutions also include savings banks and state non-member banks. They operate at smaller, regional scales and were not included in our study.} 
We use the consolidated asset total as a proxy for customer base and consumer reach: 
these banks account for 97.3\% of all assets held by FDIC-insured commercial banks (52.8\% of them by number).
We used the FDIC's BankFind Suite \cite{fdic_bulk_data_download} to obtain bank website URLs and branch locations based on their RSSD IDs. 
We removed 19 duplicate URLs associated with banks owned by the same holding company,\footnote{A bank holding company may include multiple legally distinct banks (each with a unique RSSD ID) but typically provides a shared privacy policy across subsidiaries.} 
as well as 19 duplicates where different URLs redirected to the same website due to mergers or acquisitions.
We retained only the highest-ranked occurrence in each duplicate case, resulting in 2,091 unique bank website URLs. Because our study includes analysis of banks' CCPA-related disclosures and opt-out implementations, we collected all data from a California vantage point using a commercial VPN. 
We successfully accessed 2,073 bank websites, which comprise the final dataset for this study. For each bank, we collected its (1) privacy policies, (2) cookie opt-out controls, (3) CCPA opt-out links, (4) responses to the Global Privacy Control (GPC) signal, and (5) examined third-party cookies. 
Data collection ran from October 2024 to January 2025. 
Figure ~\ref{fig:data-collection-method} shows our data collection pipeline. 

\subsubsection{Privacy policy collection}\label{sec:policy-collection}
To retrieve privacy policies, we first used a keyword-based web crawler to identify relevant links and download the corresponding documents. The downloaded notices were then cleaned and classified based on their headings.

\paragraph{Privacy policy retrieval.}
We built a custom crawler based on Scrapy~\cite{scrapy} that searches for links containing a list of privacy policy-related keywords in either the link's anchor text or URL.
Our keyword list was informed by prior research~\cite{amos_privacy_2021, libert_automated_2018, wagner_privacy_2023, srinath_privacy_2021} and iteratively refined and tested on a sample of 100 banks. 
Our crawler searched for relevant links up to a depth of three to ensure reliable discovery of privacy policies. 
We manually visited 373 bank websites where the crawler failed to access pages, found no privacy-related content, or no GLBA notices (which we expected all banks to provide).
We also manually reviewed the top 200 banks due to more complex site structures. 
In total, we collected 11,265 potential privacy policy files. Through automated and manual review, we filtered out non-relevant files and duplicates, resulting in 3,305 unique privacy files across 2,069 banks.

\paragraph{Policy classification.} 
Often banks had a webpage titled ``Privacy Policy'' that contained a collection of privacy policy information, such as a GLBA notice, a CCPA privacy policy and/or a CCPA notice at collection, data practices related to their online and mobile services, and/or a cookie policy. 
Some banks presented some or all notices in separate, stand-alone policies. 
To answer RQ1 (number of privacy policies a bank provides), 
we manually labeled files based on headings and categorized them into five types: 
(1) \textit{GLBA notice,} typically titled ``U.S. consumer privacy notice,'' or ``privacy policy.'' It is typically presented as a PDF or sometimes as HTML, either standalone or combined with other notices and resembles the GLBA short-form template.
(2) \textit{General privacy policy,} often titled ``privacy policy'' or ``digital privacy'' that focuses on online privacy, or may combine elements that would otherwise appear in the following three types of standalone notices:
(3) \textit{CCPA privacy policy,} by which we refer to both the privacy policy and the notice at collection provided in compliance with the CCPA;
(4) \textit{mobile privacy policy,} specific to mobile applications or mobile data collection practices;
(5) \textit{cookie/advertising policy,} focusing on cookie use and tracking technologies or interest-based advertising.

\subsubsection{Opt-Out control collection} 

We collected three types of opt-out controls: GLBA, CCPA, and cookie-related controls. \textit{GLBA opt-out mechanisms} are typically described within the GLBA privacy notices we collected. 
We collected CCPA and cookie opt-out controls separately through additional crawling:

\paragraph{Cookie opt-out controls} 
To identify a bank's cookie opt-out controls (if provided), we used a keyword-based web crawler similar to the one for privacy policies. 
The crawler searches for keywords related to cookie control (e.g., ``cookie,'' ``control,'' ``setting,'' ``manage'') to flag websites that might contain such links and only search for links that are on the main webpage (depth=1). We then manually visited each flagged website to verify whether it included a cookie control link. If it did, we accessed and saved the content of the corresponding cookie control page. In total, we identified 96 banks that provided cookie controls on their websites.

\paragraph{CCPA opt-out controls}
CCPA allows for opting out of the sale or sharing of personal information via an opt-out link and frictionless opt-out preference signals. We adapted Tran et al.'s~\cite{tran2024measuring} keyword-based web crawler, which detects CCPA's opt-out links by their CCPA-required labels. 
We manually verified each page and identified 45 banks that provided an opt-out link. 

To measure websites’ responses to GPC signals, we used the GPC web crawler developed by Hausladen et al.~\cite{hausladenwebsites} to send GPC signals and receive responses from each website. 
For the 234 websites the crawler could not access, we manually visited each one using a California IP and used the OptMeowt Chrome extension \cite{optmeowt} to send GPC signals and record each website’s response. During these visits, we observed that 6 websites displayed messages indicating that opt-out preference signals were honored, even though the OptMeowt extension reported no detectable GPC policy. These messages typically appeared after clicking the CCPA opt-out link. To be conservative, we treated these cases as respecting GPC signals. In total, 64 were found to respect GPC signals.

\subsubsection{Third-party cookies}\label{sec:third-party-cookies-method} 
We used third-party cookies as one indicator of whether a bank shares consumer information with external entities. 
We rendered each site using a Selenium-based web browser. We then used the Chrome DevTools Protocol’s Network domain to monitor network activity and collect all cookies stored during the browser session. Each cookie’s domain was compared to the domain of the website visited; if the domains did not match, the cookie was classified as a third-party cookie. After collecting third-party cookies from each website, we attempted to infer their purpose by comparing their domains against EasyList's known list of advertising-related trackers~\cite{easylist}. If no exact match was found, we compared only the last two segments of the cookie’s domain with the list (e.g., "unrulymedia.com" instead of ".targeting.unrulymedia.com"), which improved detecting advertising-related cookies. We found 1,454 banks (70.1\%) allowing third-party cookies on their websites, 1,252 (60.4\%) of which contained marketing cookies. 

\subsection{Privacy Policy Analysis}

\subsubsection{Plain text extraction.}\label{sec:plain-text}
To address RQ1 on the amount of privacy information provided by each bank, we extracted the plain text for all of a bank's policy files.
For PDF files, we used a vision-based LLM pipeline (GPT-4o) to extract and compile text from each page, 
as text extraction methods (e.g, using \textit{PyMuPDF}~\cite{pymupdf}) proved unreliable due to the visual complexity of the layouts (e.g., flipped cell texts in GLBA notice tables).
For HTML files, we used \textit{Boilerpipe}~\cite{boilerpy3} to extract the main text and \textit{BeautifulSoup}~\cite{beautifulsoup4} as a fallback. 
Then, we merged all of a bank's processed privacy policy files into one plain-text file.
For readability analysis, we used the widely-adopted Flesh-Kincaid Grade Level~\cite{shedlosky2009tools} that measures the difficulty of reading a text based on sentence length and word complexity with a score corresponding to a U.S. school grade. 

To address RQ2 \& 3 on marketing/advertising-related third-party sharing statements and opt-outs, we analyzed the content of privacy policies as detailed below. The analysis of opt-out links/controls is relatively straightforward and thus not discussed here.

\subsubsection{GLBA notice analysis}

GLBA notices mostly follow the standardized GLBA template, with the majority in PDF or HTML format. We analyzed the data-sharing table, the opt-out box, and the OII box in them (see Figure~\ref{fig:glba-model}). 
For PDFs, we employed our 
vision-based LLM pipeline and refined it given the known GLBA table format: we first manually annotated the page numbers of each target item and then extracted the relevant items from each page into a prescribed format using task-specific prompts. 
We iteratively refined our prompt until achieving satisfactory performance. We further manually verified and corrected all results. 
For HTML-based GLBA notices, we parsed them with \textit{BeautifulSoup} and used \textit{regex} to identify and extract the three target items, followed by manual verification and correction. 
Once extracted, we analyzed a GLBA notice's data-sharing table and opt-out box with \textit{regex} and manually annotated the OII box.

\subsubsection{Non-GLBA policy analysis}\label{sec:non-glba-analysis}

General, online, mobile, and CCPA privacy policies rarely disclose the specific data types being shared with third parties and refer to them broadly as \textit{``personal information''} or \textit{``your information''}, and occasionally specify a particular data type (e.g., medical information). In contrast, the third-party cookies disclosed in cookie policies are usually associated with data collected during a user's online activity.
Privacy policies use different language to describe third-party sharing practices for marketing. The CCPA defines ``sharing'' as the disclosure of personal information to a third party for \textit{cross-context behavioral advertising}, and some CCPA policies adopt this terminology. Other policies, including some CCPA policies, use broader phrasing (e.g., ``we work with advertising companies''). 
In both cases, the practice involves consumer information being disclosed to an entity other than the one that collected it, for use in marketing, though the language used has different regulatory implications. 

While existing automated classification methods can identify if a privacy policy segment relates to third-party sharing/collection~\cite[e.g.,][]{srinath_privacy_2021,harkous2018polisis}, they are trained on the OPP-115 corpus developed in 2016~\cite{wilson2016creation}, which pre-dates CCPA and thus lacks respective notices.
Other automated methods~\cite{andow_policylint_2019, andow_actions_2020, hooda2024policylr} extract third-party sharing statements based on pre-defined data types and entity taxonomies, which may lead to incomplete results. 
Instead, we used a bottom-up manual annotation approach to capture third-party sharing statements as written and respective nuances. 

\subsubsection{Manual annotation approach} 

The first author developed an initial codebook drawing from both inductive codes from a preliminary analysis of 60 notices and relevant codes from OPP-115~\cite{wilson2016creation}. 
The codebook was refined over multiple rounds of annotation and discussions among three co-authors on subsets of additional notices (98 in total) across various privacy policy types, structures, formats, and lengths. 
In each round, the three co-authors independently annotated a set of 5–10 policies, discussed disagreements, and clarified definitions in twice-weekly training sessions over eight weeks. After each round, the first author refined the codebook by compiling lists of common and vague phrases and identifying cross-document matches. This process continued until high inter-rater reliability was achieved (Krippendorff's Cu-$\alpha$~\cite{gonzalez-prieto_reliability_2023, krippendorff_content_2019}: 0.95 for CCPA privacy policies; 0.89 for general, mobile, and cookie policies). 

The final annotation scheme consisted of several code groups that are often applied in combination: affirmation or denial of third-party sharing/selling of personal information, data types involved, sharing purposes, opt-out choice type, and opt-out choice scope. 
We did not include the receiving entity, as this information was typically vague or not provided. 
Since privacy policies are often vague or ambiguous~\cite{reidenberg2016ambiguity},  we included an ``unclear'' option in each group. We developed distinct (but partially overlapping) codebooks for CCPA-specific and non-CCPA content due to their differing language, 
 which improved annotation accuracy and IRR. 
Annotators applied different code combinations to semantically different segments. Given the large volume of privacy policies and the extensive training our annotators underwent to achieve high inter-rater reliability, each policy was annotated once by a single annotator.\footnote{We provide the collection of documents,  codebook, and annotations in a Zenodo repository: \url{https://doi.org/10.5281/zenodo.17014519}.}

\subsubsection{Resulting dataset}

We successfully collected privacy policies from 2,069 bank websites (4 lacked policies), and analyzed third-party cookies on 2,070 websites (3 could not be rendered by our web browser). 
We classified the content of all identified privacy policies into five types defined in Section~\ref{sec:policy-collection}.
Among the 2,004 banks for which we found GLBA notices, 2 of them did not include a data-sharing table or disclose sharing practices. 14 banks provided more than one GLBA notice tailored to their different business lines, such as savings, wealth management, and home loans.\footnote{For 9 banks, the data-sharing table was consistent across their own GLBA notices. Disclosures on nonaffiliate sharing varied for 5 banks. Opt-out boxes differed only in minor details, such as the exact contact methods, but otherwise were the same for all.} 
For consistency, we analyzed one GLBA notice per bank (notice covering checking/savings services that all banks in our dataset provide). This resulted in a final sample of 2,002 GLBA notices, on which our RQ2 GLBA-related results are based. \looseness=-1

\subsection{(In)consistency Analysis}

We analyzed (in)consistencies in a bank's third-party sharing statements across its policies (RQ2), and between disclosed third-party sharing and the availability of corresponding opt-outs (RQ3).\looseness=-1

\subsubsection{Third-Party sharing statements}
To identify inconsistencies, we matched up third-party sharing statements in GLBA notices with those in general, mobile, and cookie policies, as well as CCPA statements. Notably, these different policy types use different language to describe related third-party sharing practices. 
To compare disclosures across policies,
we identified 15 types of disclosed sharing practices based on our analysis (4 GLBA-specific; 8 in general, mobile, and cookie policies; 3 CCPA-specific). 
We then analyzed (mis)matches among these sharing practices across a bank's policy documents. We discuss the sharing practice types, how they match up, and our respective (in)consistency findings in Section~\ref{sec:rq2}. 

\subsubsection{Disclosures and opt-out choices}

To answer RQ3, we examined when a third-party marketing-related sharing practice is disclosed in a privacy policy, whether the legally required opt-outs are also provided.
For GLBA, we focused on statements on sharing with affiliates and nonaffiliates for marketing purposes, for which the GLBA requires an opt-out. For CCPA, we focused on statements regarding sale/sharing practices, for which CCPA requires an opt-out.
While not legally required by U.S. law, some banks that follow GDPR-related practices also offer opt-outs for third-party cookies used for marketing/advertising and analytics/research purposes. We therefore also included them in our analysis of opt-out availability. In addition, we compared the presence of third-party advertising-related cookies on websites with the corresponding privacy statements to identify practice-to-disclosure inconsistencies.

\subsection{Limitations}

Our study may have several limitations. First, our data collection may not capture all relevant privacy policies due to variation in how banks name and structure these documents on their websites. To maximize coverage, we (1) used an inclusive list of keywords, (2) crawled sites up to three levels deep, and (3) manually inspected sites that our crawler could not access or where key notices (i.e., GLBA) or any privacy-related documents were not detected.
Second, annotators' interpretation of the privacy policies we analyzed may have led to minor inaccuracies in qualitative coding. To ensure reliability, we developed detailed guidelines, curated a reference list of agreed segments, conducted extensive training, and achieved high inter-rater reliability among three annotators.


\section{Findings}

\subsection{RQ1: Amount of Information}\label{sec:rq1}

\begin{table}[t]
    \caption{Privacy policies provided by banks (n=2,069). Row sums indicate how many banks provide each combination; column sums show totals for each policy type.}
    \centering
    \scalebox{0.7}{
    \begin{tabular}{p{2.2cm}|lllll|rr}
    \hline
    \toprule
        & \multicolumn{5}{c|}{\textbf{Privacy Policy Type}}  &  &   \\ 
        \cline{2-6} 
        \textbf{Category}& \rotatebox{90}{\textbf{GLBA}}  &  \rotatebox{90}{\textbf{General}} & \rotatebox{90}{\textbf{Mobile}} & \rotatebox{90}{\textbf{CCPA}} & \rotatebox{90}{\textbf{Cookie}} &  \textbf{Count} & \textbf{Total}  \\ 
        
        \toprule
        GLBA only& \filledC &  \emptyC & \emptyC & \emptyC & \emptyC &  1,089 & 1,089  \\ 
        \cline{1-8} 
         GLBA+1 other & \filledC  &  \filledC  & \emptyC & \emptyC& \emptyC & 407  & 653  \\ 
        & \filledC &  \emptyC & \filledC & \emptyC  & \emptyC & 167  &   \\ 
        & \filledC &  \emptyC & \emptyC & \filledC  & \emptyC  & 58  &   \\ 
        &  \filledC &  \emptyC & \emptyC & \emptyC  & \filledC &  21 &   \\ 
        \cline{1-8} 
        GLBA+2 others& \filledC & \filledC  & \emptyC & \filledC  & \emptyC  & 109  & 221  \\ 
        & \filledC & \filledC  & \filledC  & \emptyC  & \emptyC &  78 &   \\ 
        & \filledC &  \filledC  & \emptyC & \emptyC  & \filledC &  14 &   \\ 
        & \filledC &  \emptyC & \filledC  & \filledC  & \emptyC   & 13  &   \\
        & \filledC & \multicolumn{4}{c|}{Other combinations} & 7 &   \\ 
        \cline{1-8} 
        GLBA+3/4 others& \filledC &  \filledC  & \filledC  & \filledC  & \emptyC   & 28  & 41  \\ 
        & \filledC &  \filledC  & \emptyC  & \filledC  & \filledC   & 10  &   \\ 
        & \filledC & \multicolumn{4}{c|}{Other combinations} & 3 &   \\ 
        \hline
        No GLBA, 1 other& \emptyC & \filledC & \emptyC & \emptyC & \emptyC  & 41 &  44 \\ 
        & \emptyC & \multicolumn{4}{c|}{Other combinations} & 3 &   \\ 
        \cline{1-8} 
        No GLBA, 2/3/4 others& \emptyC & \multicolumn{4}{c|}{Other combinations} & 21 & 21  \\ \hline
        Sum&2,004 &709 & 305 & 233 & 61 &   & 2,069 \\  
        \bottomrule
    \end{tabular}
    }
    \label{tab:doc-count}
\end{table}

Table~\ref{tab:doc-count} summarizes the types of privacy policies provided by each bank and the number of banks offering each combination. We found GLBA notices for 2,004 banks, and 45.2\% of all banks provided multiple privacy policies.
915 banks (44.2\%) provided one or more additional privacy policies besides their GLBA notice; the most common addition being a general privacy policy (407, 19.6\%), followed by a mobile privacy policy (167, 8.1\%).
Among banks that provided two or more additional privacy policies besides GLBA notices (262, 12.7\%), the most common combination (109, 5.3\%) included a GLBA notice, a general privacy policy, and a CCPA-specific notice. This variety demonstrates that for many banks, their privacy disclosures are layered and potentially fragmented across multiple privacy policies. \looseness=-1

We used word count and readability as proxies for the amount and complexity of information that a bank's privacy policies present to consumers. When combining all privacy policies from each bank, total word counts varied widely across banks, with most falling between 554 and 2,192 words ($median$=769, $mean$=1,572). The much higher mean compared to the median reflects a right-skewed distribution driven by a subset of exceptionally lengthy policies (e.g., privacy policy embedded in a large PDF file that also includes statements like online banking agreements). Their readability as measured by FKGL falls mostly within the high school to early college range ($median$=13.2, $mean$=13.6). Given that FKGL scores correspond to U.S. school grade levels, this indicates that most banks' content requires reading skills well above the average U.S. adult reading level of 8th grade~\cite{USreading}.

\begin{figure}[t]
    \centering
    \includegraphics[width=0.95\columnwidth]{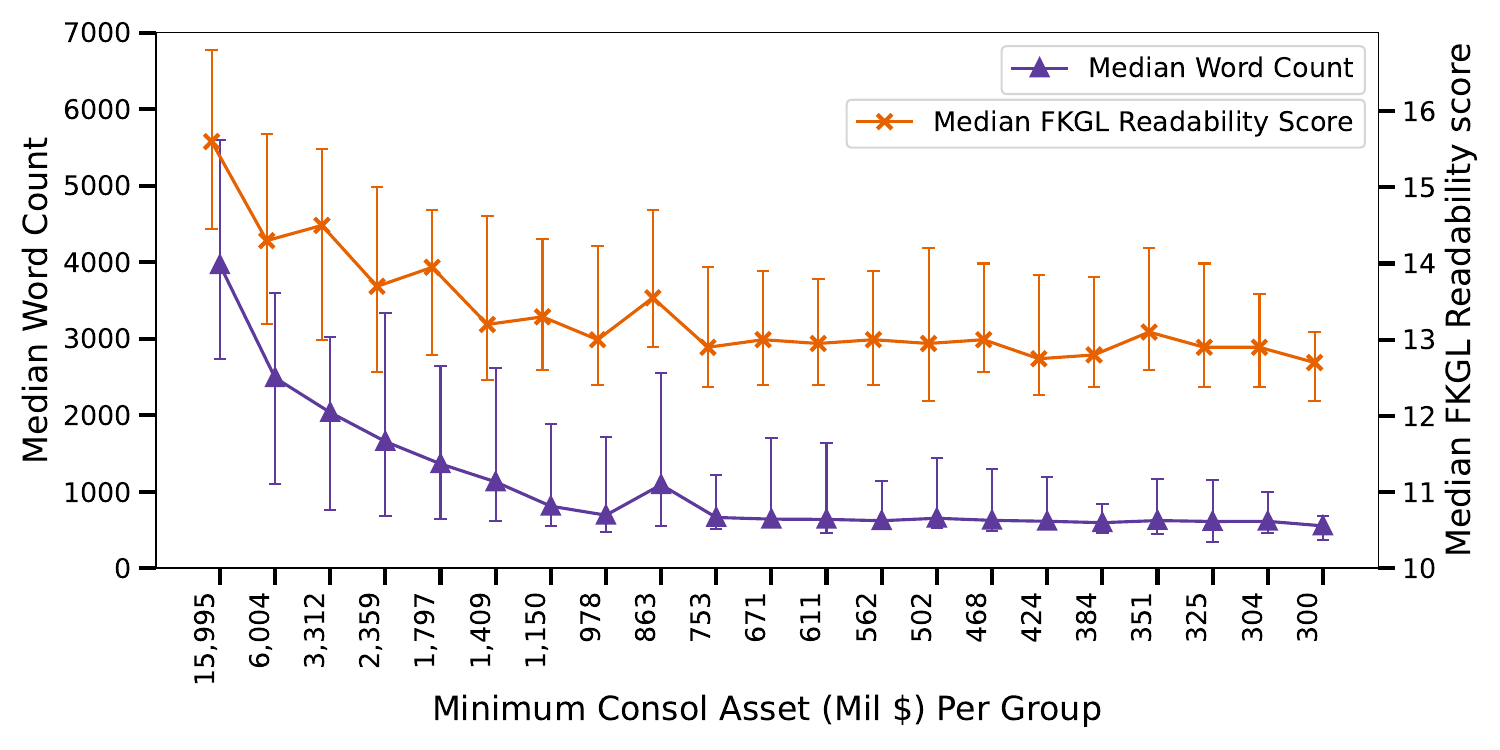}  
    \caption{Word count and readability of all policies combined per bank. Banks are ranked by consolidated assets, a proxy for their sizes, and are then grouped into sets of 100 by rank for visualization purposes. Larger banks tend to provide more privacy policy content, but their policies are less readable.}
    \label{fig:wc-readability}
\end{figure}

We used bank rank by consolidated assets as a proxy for bank size, where a lower rank indicates a larger bank.
We found a modest but statistically significant negative correlation between bank rank and word count ($\rho$=-0.38, $p$$\ll$0.001) and between bank rank and readability score ($\rho$=-0.28, $p$$\ll$0.001), as shown in Figure~\ref{fig:wc-readability}. 
The largest-sized (i.e., the lowest-ranked banks with consolidated assets $\geq$\$15,995 million) provided significantly more privacy policy content ($median$=4,017) in contrast to smaller banks offering content generally below a 1,500 word count. Larger banks also tend to use more complex language, with FKGL readability scores above 15, and smaller banks average closer to 13 or below.  

\textbf{RQ1 summary.} Both the types of privacy policies and the amount of information vary widely across banks. About half of all banks (936, 45.2\%) provided at least two privacy policies, and most banks' privacy policies are difficult to read.
Larger banks provide more privacy policy content with more difficult language compared to smaller banks.
These results show that despite the GLBA short notice’s intent to simplify privacy disclosures, banks now surround it with multiple, lengthy policies (general, CCPA, cookie), subverting the GLBA notice’s purpose and creating readability/comprehension issues that related work has documented in other sectors. \looseness=-1

\subsection{RQ2: Third-Party Sharing Statements}\label{sec:rq2}

Regarding inconsistencies in marketing-related sharing statements across policies, we first discuss the GLBA statements (\ref{sec:glba-data-sharing}). After presenting relevant statement categories we identified in other policies (\ref{sec:statement-categories}), we compare banks' denying GLBA statements (\ref{sec:no-yes}) and affirmative GLBA statements (\ref{sec:yes-no}), each with opposite statements in other polices. Last, we point out common misleading language use for no-sharing (\ref{sec:permitted-required}).

\subsubsection{GLBA third-party sharing statements}\label{sec:glba-data-sharing}

Of the seven sharing purposes prescribed in the GLBA model notice's data-sharing table (see Figure~\ref{fig:glba-model}(a)), we focused on the four relevant to third-party sharing:
1,542 banks (77\% out of 2,002 banks that provided this row) stated they are sharing consumer personal information for marketing (\textit{marketing}), 
879 (43.9\% out of 2,002) disclosed sharing with nonaffiliated financial companies for joint marketing (\textit{joint marketing}), 
329 (16.4\% out of 1,070) disclosed sharing with affiliates for affiliates' marketing (\textit{affiliate sharing}),
and 65 (3.2\% out of 1,925) disclosed sharing with nonaffiliates for nonaffiliates' marketing (\textit{nonaffiliate sharing}).
A few banks (36, 1.8\% out of 2,002 banks with a GLBA notice) shared data for all four marketing purposes, 10.6\% (214) for three of them; a third for two purposes (643, 32.1\%) and another third shared for one of these purposes (743, 37.1\%). Only 5.6\% (112) indicated not sharing for any of the four purposes. 
These four GLBA-defined third-party sharing practices have nuanced differences, 
however, for consumers, they all describe scenarios in which their personal financial information is shared by their bank with other entities for marketing use. 

Some banks include additional data-sharing disclosures in their GLBA notices' OII box. Among the 718 banks that had a non-empty OII box, 332 included California-specific language. 300 of them provided statements further limiting data-sharing practices for California residents. The most common phrasing was:
\textit{``
We will not share personal information with non-affiliates either for them to market to you or for joint marketing without your authorization. We will also limit our sharing of personal information about you with our affiliates to comply with all California privacy laws that apply to us.''}---which suggests that without consent, the bank does not share for joint marketing or nonaffiliate marketing for Californians, though remaining vague on affiliate sharing.
20 (of 332) banks included minimal or opaque disclosures that mentioned CalFIPA but only stated \textit{``In accordance with California law, [bank] does not share personal information we collect except as permitted by law.''}
55 (of 332) mentioned or linked to the bank's CCPA policy (e.g., \textit{``Please visit our California Privacy Rights Act Policy for more information.''}). These references at least acknowledge the existence of additional privacy policies, but still require consumers to navigate beyond the standardized data-sharing table in the GLBA notice, into a less prominent section of the notice, follow a link, and then read those additional policies to fully understand the bank's data-sharing practices.\looseness=-1

\subsubsection{Third-party sharing statements in other privacy policy types}\label{sec:statement-categories}

For a bank's other (non-GLBA) privacy policies, we focused on what kind of third-party sharing for marketing practices are described and which data types are referenced in these statements. 
We found that the phrasing of third-party sharing disclosures rarely reflected legal distinctions in a way that would be meaningful or recognizable to consumers. For example, ``personal information'' is used in data-sharing statements across GLBA, general, mobile, and CCPA policies, yet the associated meaning differs based on definitions given in other places in those documents. We captured data types just as they are presented, since consumers are unlikely to recognize and distinguish associated legal nuances~\cite{chen2021fighting}.
Similarly, banks' general privacy policies often used broad terms like ``partners,'' ``service providers,'' or ``contractors'' without explicitly defining them as external, third parties or naming specific entities. 
However, these terms all suggest third-party relationships. 

Thus, for the purpose of comparing across policies, we created statement categories based on whether a third-party sharing statement refers to personal information in general or to a specific data type, and whether it indicates sharing with external entities for marketing-related purposes.
These categories include both statements allowing/describing third-party sharing (``yes'') and explicitly denying sharing (``no''). However, in contrast to GLBA disclosures that require explicit ``no'' statements, most banks' policies are silent on sharing practices that the bank presumably does not engage in.\footnote{See examples of the sharing statement types in our Zenodo repository:  \url{https://doi.org/10.5281/zenodo.17014519}} 
We treat marketing and advertising as similar purposes, hereafter referred to as marketing.

\textit{General, mobile, and cookie policies} typically did not contain statements exactly matching the four GLBA sharing purposes; instead they fell into seven categories:
(1) Sharing personal information for marketing purposes (yes: 170 banks; no: 50 banks);
(2) Sharing specific data types (e.g., personal information provided through email, medical information) for marketing purposes (yes: 75; no: 60). 
Although our primary focus is on marketing and advertising, many of these policies also refer to analytics or research purposes (hereafter referred to as analytics) in third-party sharing statements, which is vague and may include uses related to marketing: 
(3) Sharing personal information for analytics purposes (yes: 105; no: n/a).

In addition, many banks' policies mentioned allowing third parties to place cookies or other tracking technologies on their websites:
(4) Allowing third-party marketing cookies (yes: 225; no: 4), and
(5) Allowing third-party analytics cookies (yes: 379; no: n/a).
Some policies contained vague, catch-all statements for data sharing practices:
(6) No sharing unless required by law (64 banks), and
(7) No sharing unless permitted by law (115 banks). The latter is particularly concerning as it suggests that ``no sharing'' is the bank's default, whereas the bank may actually be sharing to the fullest extent legally possible.
Some policies contained a statement on sale practices:
(8) Sale of personal information (yes: 6; no: 145).

In \textit{CCPA-related} privacy policy content, we identified two kinds of sharing statements. Some specifically referred to the CCPA definition of sharing or sale of personal information:
(1) Sharing as defined by CCPA (yes: 24; no: 105),
(2) Sale as defined by CCPA (yes: 24; no: 217).
Others also described sharing practices in CCPA disclosures, but it is unclear whether it falls strictly under the definition of CCPA:
(3) Sharing for marketing purposes, without referencing CCPA definitions (89 banks).

\subsubsection{Third-party sharing despite negative GLBA statements.}\label{sec:no-yes}

\begin{table}[]
\caption{Number of banks that indicated a ``no'' to GLBA sharing purposes (rows) but ``yes'' to related sharing categories (columns). The final column shows the count of banks with at least one inconsistency.}
\label{tab:no-yes-5}
\resizebox{\columnwidth}{!}{%
\begin{tabular}{l|rrr|rr|rr|r}

\toprule
\multicolumn{1}{c|}{\multirow{2}{*}{\textbf{{[}No{]}\textbackslash[Yes{]}}}} & \multicolumn{3}{c|}{\textbf{Per info}}       & \multicolumn{2}{c|}{\textbf{3\textsuperscript{rd} party cookies}} & \multicolumn{2}{c|}{\textbf{CCPA}} & \multirow{2}{*}{\textbf{Count}} \\ \cline{2-8}
\multicolumn{1}{c|}{}                                           & \textbf{Mkt} & \textbf{Anlt} & \textbf{Spec} & \textbf{Mkt}         & \textbf{Anlt}         & \textbf{Shar}    & \textbf{Sale}   &                              \\ \hline
\textbf{Mkt}                                                    & 7            & 6             & 8             & 14                   & 52                    & 5                & 0               & 68                           \\
\textbf{Joint-mkt}                                              & 56           & 46            & 22            & 86                   & 178                   & 42               & 6               & 257                          \\
\textbf{Aff-mkt}                                                & 44           & 27            & 19            & 66                   & 110                   & 40               & 6               & 172                          \\
\textbf{Nonaff-mkt}                                             & 136          & 89            & 66            & 188                  & 321                   & 96               & 18              & 481                          \\
\bottomrule
\end{tabular}%
}
\end{table}

We assessed when banks indicated a ``no'' for each of four GLBA marketing-related sharing purposes (marketing, joint marketing, affiliate sharing, nonaffiliate sharing), how many banks simultaneously disclosed a ``yes'' to third-party sharing in related categories identified above. 
We found that many banks 
still disclosed varying amounts of related sharing in other policies despite negative indications in their GLBA notice (see Table~\ref{tab:no-yes-5}). 

\textit{Inconsistencies regarding nonaffiliate sharing.}
1,860 banks (92.9\% of 2,002 banks with an analyzable GLBA notice) stated in their GLBA notice that they do not share with nonaffiliates. Yet 481 of them (26.6\%) also had affirmative sharing statements in other policies under at least one category we identified. 
Among these 481 banks, about 30\% stated sharing personal information for advertising purposes, 
more than two-thirds disclosed using third-party analytics cookies, 
and about 40\% for third-party marketing cookies. 
About 20\% disclosed sharing personal information for marketing in CCPA-related disclosures. 
Possibly less concerning but still an indication of nonaffiliate sharing, 
less than 20\% stated sharing for analytics purposes. 
More than 10\% disclosed sharing specific personal information types for marketing (e.g., \textit{``[w]e may also share your device's physical location, combined with information about what advertisements you viewed and other information we collect, with our marketing partners [...]''}). 
A small percent, but still 17 banks in number, 
even disclosed selling data under CCPA (e.g., \textit{``In the preceding twelve (12) months, we have sold personal information''}).

These inconsistencies may reflect GLBA's narrow scope of personal information as relating to financial information only, whereas other privacy policies often cover a broader range of practices, including data collected through a bank’s website and mobile apps.
Under GLBA, a ``no'' to nonaffiliate sharing means that the bank does not share (financial) personal information with nonaffiliates for \textit{those nonaffiliates'} marketing purposes. However, this does \textit{not} preclude sharing with nonaffiliates for other purposes like supporting the bank’s own marketing efforts.
This discrepancy is concerning as it may mislead consumers. A consumer may conclude from the GLBA notice that a bank does not share their personal information with others, while in practice (with a 25.8\% likelihood) the bank still shares lots of their personal information with third parties.

\textit{Inconsistencies regarding affiliate sharing.}
A similar definitional gap exists for affiliate sharing, which under GLBA is confined to sharing personal information with affiliated companies (e.g., a bank's affiliated investment or loans business) for \textit{their} marketing. Among the 741 banks that stated they do not share for affiliate marketing in their GLBA notice, 
23.2\% of them (172 banks) still disclosed other data-sharing practices in other policies. More than two-thirds of these 172 banks disclosed that they use third-party analytics cookies, more than a third disclosed third-party marketing cookies, and about a quarter stated sharing for marketing.

\textit{Inconsistencies regarding other sharing.}
We have seen above that when inconsistencies occur, most of them are between GLBA and third-party cookie disclosures. This pattern persists for ``no'' responses to GLBA-defined joint marketing and general marketing. 
Among the 1,123 banks that said they do not share for joint marketing in GLBA notice, 257 banks (22.9\% of them) in fact stated that they do share personal information for marketing purposes in other privacy policies. For those with at least one inconsistency, about 70\% banks stated allowing third-party analytics cookies, and a third for marketing cookies. About a quarter of them stated sharing personal information for marketing.
A smaller group of 460 banks indicated ``no'' to sharing for the GLBA-defined marketing purpose, and 68 (14.8\% of them) disclosed at least one sharing practice in other policies. 
Compared to other purposes, under joint marketing and marketing, we found smaller percentages of banks that had at least one inconsistency.
Yet, any inconsistencies under these two purposes are striking, since they are about sharing for the bank’s own marketing purposes, a practice likely more common than nonaffiliate or affiliate sharing that is more narrowly defined under GLBA.\looseness=-1

Taken together, of the banks that indicated ``no'' to at least one GLBA-defined sharing purpose, 492 also disclosed at least one related data sharing practice as occurring in other policies. They account for 53.8\% of the 915 banks that provided a GLBA notice and at least one additional policy. 
This suggests that even when banks claim not to share under GLBA categories, they are very likely to still engage in similar data-sharing practices for marketing.
In sum, inconsistency between GLBA's and other policies' disclosures is widespread, which highlights coverage gaps and the lack of compatible clarity across privacy disclosures.

\subsubsection{More restrictive third-party sharing than affirmative GLBA statements}\label{sec:yes-no}

{\small
\begin{table}[]
\caption{Number of banks that indicated a ``yes'' to GLBA sharing purposes (rows) but ``no'' to related sharing categories (columns). The final column shows the count of banks with at least one inconsistency.}
\label{tab:yes-no-4}
\begin{tabular}{l|rrr|rr|r}

\toprule
\multicolumn{1}{c|}{\multirow{2}{*}{\textbf{[Yes]\textbackslash{}[No]}}} & \multicolumn{3}{c|}{\textbf{Per info}}  & \multicolumn{2}{c|}{\textbf{CCPA}} & \multirow{2}{*}{\textbf{Count}} \\ \cline{2-6}
\multicolumn{1}{c|}{}                                                & \textbf{Mkt} & \textbf{Spec} & \textbf{Sale} & \textbf{Shar}    & \textbf{Sale}   &                                   \\ \hline
\textbf{Mkt}                                                         & 37           & 41            & 114           & 184              & 88              & 326                               \\
\textbf{Joint-mkt}                                                   & 19           & 30            & 65            & 96               & 52              & 192                               \\
\textbf{Aff-mkt}                                                     & 7            & 14            & 20            & 84               & 36              & 117                               \\
\textbf{Nonaff-mkt}                                                  & 1            & 4             & 6             & 13               & 4               & 21                                \\
\bottomrule
\end{tabular}%
\end{table}
}

Some banks that indicated they share personal information under GLBA-defined categories simultaneously stated in their other privacy policies that they do not share information with third parties, most commonly in their CCPA-related sharing disclosures (see Table~\ref{tab:yes-no-4}).  

Among the 329 banks that disclosed affiliate sharing in their GLBA notice (16.4\% of 2,002 banks with an analyzable GLBA notice), 117 banks (35.6\%) had at least one statement denying data-sharing in their other policies. 71.8\% these 117 banks stated that they do not share personal information, with or without reference to CCPA's definition of sharing in CCPA policies or California-specific sections, and 30.8\% of them stated that they don't sell personal information in CCPA-related disclosures.

Regarding the GLBA's nonaffiliate sharing, similarly high percentages of banks disclosed at least one denying data-sharing statement in their other policies, though fewer banks in number did so. Among the 65 banks (3.2\% of 2,002 banks with an analyzable GLBA notice) that stated a ``yes'' to nonaffiliate sharing, 30.1\% of them (21 banks) disclosed a denial of data-sharing of some kind. 62\% of the 21 banks indicated that they do not share, and 19\% do not sell, in CCPA-related disclosures.

Among banks that indicated a ``yes'' to GLBA's marketing or joint-marketing purposes, smaller percentages (about 20\% for each), though with larger counts, denied respective data-sharing elsewhere. For both, more than half of the ones with inconsistencies disclosed not sharing or selling in CCPA-related disclosures. 

In total, 337 banks that disclosed sharing under GLBA indicated in other policies that they do not share/sell in CCPA-related sections. 
This pattern of discrepancy between GLBA and CCPA-related disclosures may indicate that banks specifically restrict their third-party sharing practices for their Californian customers while sharing their other customers' data more freely.

\subsubsection{Misleading ``we don't share'' statements}\label{sec:permitted-required}
We found that many ``we don't share'' statements are crafted to sound reassuring but remain vague and are ultimately misleading.
A common pattern involves using broad legal qualifiers. 
115 banks used the phrase ``except as permitted by law'' (or synonyms like ``permissible'' and ``authorized'') in their no-sharing statements, 
(e.g., \textit{``We do not disclose any nonpublic personal information about our customers or former customers to anyone, except as permitted by law.''}). 
Despite this seemingly restrictive language, many of these banks also had affirmative data-sharing statements in their privacy policies. Among the 115 banks that used ``except as permitted by law,'' 83.5\% of them (96 banks) indicated that they share personal information under at least one of the four GLBA marketing-related sharing purposes. 40\% (46 banks) stated they allow third-party marketing or analytics cookies on their websites, and 21.7\% (25 banks) disclosed sharing anonymized data. This language that suggests banks only share data when legally allowed may in fact enable extensive sharing. 

An even more restrictive-sounding qualifier ``except as required by law'' (or ``compelled'') appeared in 64 banks' no-sharing statements (e.g. \textit{``All information acquired through orders are kept confidential and will not be disclosed to third parties except as may be required by law.''}). While this language appears more protective of consumers than the broader ``except as permitted by law'' qualifier, similarly high proportions of those banks disclosed third-party sharing practices. Over 80\% of the 64 banks indicated sharing under at least one of the four GLBA purposes. 42\% stated that they allow third-party marketing or analytics cookies, and 42\% disclosed sharing anonymized data. 

In total, 137 banks used at least one of the two phrases.
These qualifiers allow banks to claim that they do not share personal information while still leaving the door open to a wide range of sharing practices. Given how starkly the no-sharing statements with the qualifiers contrast with the banks’ own data-sharing disclosures elsewhere in their policies, these qualifiers seem to function more as blanket disclaimers than meaningful disclosures.

\begin{figure}[t]
    \centering
    \includegraphics[width=\columnwidth]{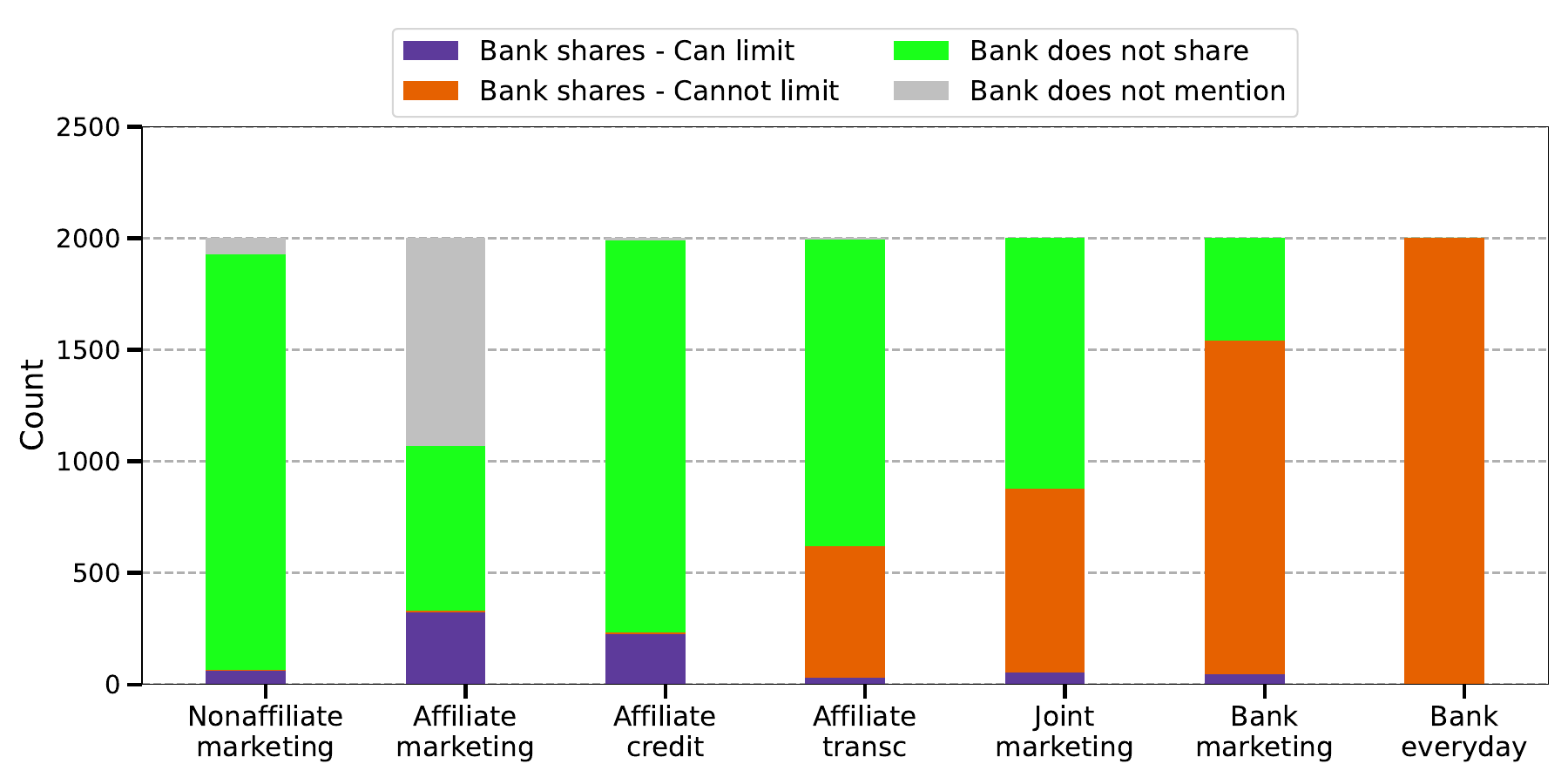}  
    \caption{Data-sharing statements and indication of whether consumers can limit the corresponding sharing practice in GLBA notices (n=2,002).}
    \label{fig:glba-disclosure}
\end{figure}

\textbf{RQ2 summary.} 
Banks' affirmative sharing disclosures under the four GLBA marketing-related categories ranged widely, from 77\% to just 3\%.
Comparing these with disclosures in other privacy policies, we noted two prevalent yet nuanced inconsistency types:  
(1) 492 banks (53.8\% of those with both a GLBA and other policies) said ``no'' under GLBA while saying ``yes'' elsewhere for related sharing. This highlights that GLBA ``no'' statements alone are insufficient for understanding banks’ data practices, especially when online and mobile services are involved. 
(2) In contrast, 337 banks (36.8\% of those with both a GLBA and other policies) disclosed sharing under GLBA and yet restricted such practices specifically for California residents. 
Additionally, although not contributing to the inconsistency types identified above, many ``we don't share'' statements involved broad legal qualifiers like ``except as permitted by law,'' which complicates the interpretation of such no-sharing disclosures and may mislead consumers. 

\subsection{RQ3: Third-Party Sharing Opt-Outs}

\begin{table}[t]
    \centering
    \caption{Summary of opt-out controls provided (n=2,073).}
    \resizebox{0.37\textwidth}{!}{ 
    \begin{tabular}{c|ccc|rr}
    \toprule
        \textbf{\#Controls} & \textbf{GLBA} & \textbf{Cookie control} & \textbf{CCPA} & \textbf{Count} & \textbf{Total}  \\ \hline
        0 & \emptyC & \emptyC & \emptyC & 1,620 & 1,620 \\ \hline
        1 & \filledC & \emptyC & \emptyC & 311 & 378\\ 
         & \emptyC & \filledC & \emptyC & 45 & \\ 
         & \emptyC & \emptyC & \filledC & 22 & \\ \hline
        2 & \filledC & \filledC & \emptyC & 16 & 59\\ 
         & \filledC & \emptyC & \filledC & 24 & \\ 
         & \emptyC & \filledC & \filledC & 19 & \\ \hline
        3 & \filledC & \filledC & \filledC & 16 & 16\\ 
        \hline
        Sum & 367 & 96 & 81 & & 2,073\\ \bottomrule

    \end{tabular}\label{tab:all-optout}}
    
\end{table}
To evaluate whether banks implement required opt-outs, we compare their data-sharing disclosures with the actual opt-outs they implement. We first summarize the number of opt-out options offered to consumers (\ref{sec:all-optouts}), then examine sharing disclosures and opt-out implementations for GLBA (\ref{sec:glba-optout}), cookie control (\ref{sec:3-cookies}), cookie practice (\ref{sec:3-practice}), and CCPA (\ref{sec:ccpa-optout}).

\subsubsection{Third-party sharing opt-outs provided by banks}\label{sec:all-optouts}

1,620 banks (78.1\%) did not offer any sharing-related opt-outs, while 453 banks (21.9\%) provided at least one (see Table \ref{tab:all-optout}). Among these, 75 banks offered two or more opt-outs. GLBA-related opt-outs were most common (367 banks, 17.7\%), and a substantial number of banks (142, 6.8\%) provided other types: cookie-related controls (96, 4.6\%) and CCPA opt-outs (81, 3.9\%). This indicates that many banks are adopting multiple sharing opt-out types to comply with different privacy laws. It also underlines that banks' data-sharing practices often extend beyond what is covered by GLBA. 

\subsubsection{GLBA-required opt-outs}\label{sec:glba-optout}

\begin{table}[t]
    \caption{Combinations of GLBA opt-out methods provided by banks (n=371). Row sums show the number of banks offering each combination, and column sums show the total number of banks that provide a respective opt-out.}
    \centering
    \scalebox{0.72}{
    \begin{tabular}{c|ccccc|rr}
    \hline
    \toprule
        \multirow{2}{*}{\parbox{1.5cm}{\textbf{\#Opt-out methods}}} & \multicolumn{5}{c|}{\textbf{Opt-out method}} &  &   \\ 
        \cline{2-6} 
         &  \rotatebox{90}{\textbf{Phone}} & \rotatebox{90}{\textbf{Link}} & \rotatebox{90}{\textbf{Mail}} & \rotatebox{90}{\textbf{Email}} &  \rotatebox{90}{\textbf{Branch}} & \textbf{Count} & \textbf{Total}  \\ 
        
        \toprule
        0 &  \emptyC & \emptyC & \emptyC & \emptyC & \emptyC &  4 & 4  \\ 
        \cline{1-8} 
        1 &  \filledC  & \emptyC & \emptyC& \emptyC & \emptyC &114  & 165  \\ 
         &  \emptyC & \emptyC & \filledC & \emptyC  & \emptyC & 43  &   \\ 
         & \multicolumn{5}{c|}{Other combinations} & 8 &   \\ 
        \cline{1-8} 
        2  & \filledC & \filledC  & \emptyC & \emptyC  & \emptyC  & 96  &  156 \\ 
         & \filledC  &\emptyC  & \filledC  & \emptyC  & \emptyC &  35 &   \\ 
         &  \filledC  & \emptyC & \emptyC  & \filledC &\emptyC  &  13 &   \\ 
         & \multicolumn{5}{c|}{Other combinations} & 12 &   \\ 
        \cline{1-8} 
        3 &  \filledC  & \filledC  & \filledC  & \emptyC   &\emptyC   & 23  &  42 \\ 
         & \multicolumn{5}{c|}{Other combinations} & 19 &   \\ 
        \cline{1-8} 
        4 & \multicolumn{5}{c|}{Other combinations} & 4 & 4  \\ 
        \hline
        Sum & 307& 139& 122&33&18& &371\\
        \bottomrule
    \end{tabular}
    }
    \label{tab:glba-method}
\end{table}

Under the GLBA, banks that share for nonaffiliate marketing, affiliate marketing, or affiliate credit-related purposes must provide respective opt-outs. We found that almost all banks followed these requirements and provided required GLBA opt-outs (see Figure \ref{fig:glba-disclosure}). For sharing purposes without required opt-outs, few banks that disclosed sharing provided opt-outs (joint marketing (6.0\%), affiliate transactions (4.5\%), bank's own marketing (2.9\%), and everyday business (0.1\%)). 
This indicates that banks generally only offer opt-out controls when mandated by law. There is also a risk that banks may categorize their data sharing practices as ``joint marketing'' rather than sharing for affiliate or nonaffiliate marketing to sidestep GLBA opt-out obligations.\looseness=-1

Furthermore, we found that banks' GLBA-related opt-out options are often burdensome for consumers to exercise. Calling the bank was the most commonly mentioned opt-out method (307 of 367 banks providing GLBA-related opt-outs, 83.7\%), and for many, the only method provided (114, 31.1\% of 367). Additionally, 43 banks (11.7\% of 367) required consumers to submit opt-out requests by mail, with no other methods offered. Only a third of banks (139, 37.9\% of 367) offered a link to their webpage. 
Thus, while most banks technically comply with GLBA opt-out requirements, the opt-out methods are often unnecessarily cumbersome, potentially discouraging consumers from exercising their privacy rights.

\subsubsection{Third-party cookie controls}\label{sec:3-cookies} 

300 banks' privacy policies mentioned that consumers can opt out of their use of cookies, but through browsers, e.g., \textit{``You can set your browser to refuse Cookies.''} 106 banks in policies referred consumers to cookie opt-outs offered by third parties (e.g., the Network Advertising Initiative), often for advertising cookies specifically. Only 17 banks mentioned using their cookie setting/banner as an opt-out choice in policies.

We examined cookie controls implemented/provided on websites.
Among the 332 banks that indicated allowing third-party cookies for marketing, analytics, or both purposes in policies, only 43 (13.0\%) of them implemented cookie controls (see Table~\ref{tab:thirdpartycookies}).  
On the other hand, among the vast majority of banks (1,741) that did not disclose cookie practices in privacy policies, 53 (3.0\% of 1,741) still implemented cookie controls. 
This suggests that they are engaging in cookie-related data sharing, which potentially involves third-party sharing, despite not explicitly disclosing it. 

The 96 cookie control links we identified were referred to by 25 different names, echoing prior work that documented similarly diverse cookie consent interfaces~\cite{degeling2018we,utz2019informed}. We also found 38 distinct labels for cookies that users can opt out of, and 13 labels for cookies that users cannot. Banks' cookie controls inconsistently applied 10 of these labels: for example, 6 banks categorized ``functional'' cookies as non-opt-out-able, while 22 banks allowed users to opt out. This inconsistent labeling may contribute to user confusion, as prior work has shown that users often mistake ``functional'' cookies for ``strictly necessary'' ones~\cite{habib2022okay}. 
Taken together, the lack of opt-out availability and standardization in cookie labeling undermines transparency and may mislead consumers about their choices.

\begin{table}[t]
\caption{Disclosure about allowing third-party cookies, the availability of cookie controls, and third-party cookie practices (n=2,073).}
\label{tab:my-table}
\resizebox{\columnwidth}{!}{%
\begin{tabular}{lrr|rr}
\toprule
 &  &  & \multicolumn{2}{c}{\textbf{3\textsuperscript{rd} party cookies found} } \\ 
 \cline{4-5}

 \textbf{Sharing Disclosure} & \textbf{Total} &\textbf{Cookie control} & \textbf{Any} & \textbf{Marketing} \\ \midrule
Do not share & 0  &  0                         & 0 & 0  \\ 
For marketing   &   225     & 35       &     194   & 184 \\
Other sharing  & 107 &   8             & 86 & 75                     \\
Not available & 1,741 & 53 & 1,174 & 993\\ \hline
Sum & 2,073 & 96 & 1,454 & 1,252\\ 
\bottomrule
\end{tabular}%
}\label{tab:thirdpartycookies}

\end{table}
\subsubsection{Third-party cookie practices}\label{sec:3-practice}

Our cookie script analysis revealed that 1,454 banks' websites (70.1\%) contained third-party cookies, and 1,252 of them (60.4\%) specifically included third-party marketing cookies (see Table~\ref{tab:thirdpartycookies}). However, among the 1,252 banks with third-party marketing cookies, only 184 websites (14.7\%) disclosed any practice of allowing marketing cookies. Interestingly, we also found that of the 225 banks that disclosed allowing marketing third-party cookies, 41 (18.2\%) did not actually have such third-party cookies on their website. 
This discrepancy may stem from vague or incomplete disclosures. Banks often fail to specify who receives cookie data, making it difficult for consumers to determine the true scope of their sharing practices or whether third parties are involved in marketing activities. \looseness=-1

\begin{table}[t]
\caption{CCPA disclosure and opt-out methods (n=2,073).}
\resizebox{\columnwidth}{!}{%
\begin{tabular}{lrrrr}
\toprule
&    &\multicolumn{3}{c}{\textbf{Control type}} \\ 
\cline{3-5}
\textbf{Sell/Share Disclosure} & \textbf{Total} & \textbf{Opt-out link} & \textbf{GPC} & \textbf{Either type} \\ 
\midrule
Do not sell/share    &    200   & 18             & 21 &            31              \\
Sell/share under CCPA  & 45   & 13    &         22            &   23               \\
Other sharing & 9       & 4 &  4                 &  6    \\ 
Not available & 1,819   & 10 & 17 & 21\\ \hline
Sum &2,073 & 45 & 64 & 81\\
\bottomrule
\end{tabular}%
}\label{tab:ccpa}

\end{table}
\subsubsection{CCPA privacy controls}\label{sec:ccpa-optout}

We assessed whether the banks that are required to provide CCPA do-not-sell/share opt-out based on their sharing disclosures provided an opt-out link and responded to GPC signals (see Table \ref{tab:ccpa}). 
200 banks explicitly stated that they do not sell/share personal information under CCPA. Interestingly, several of these still provided an opt-out link (18) and respected  GPC signals (21). 
Either these banks are taking extra precautions to avoid being perceived as non-compliant with the CCPA, or their explicit assertion not to share/sell may not reflect their actual practices.
Of the 45 banks that did acknowledge to sell/share data under CCPA, 22 (48.9\%) failed to implement required opt-outs. Only 13 (28.9\%) provided an opt-out link, and 22 (48.9\%) respected GPC signals.
We also found a few banks that made no statements about selling/sharing under CCPA in their policies, yet still provided a do-not-sell/share opt-out link (10) or respected GPC signals (17). This inconsistency raises concerns that these banks may be engaging in CCPA-covered data sharing without transparently disclosing such practices in their privacy policies, or may be implementing opt-outs they are not required to provide.\looseness=-1

\textbf{RQ3 summary.}
Although banks mostly implemented the required opt-outs for GLBA, half of the banks that disclosed the sale/sharing of personal information under CCPA failed to implement the required opt-outs. Sixty percent of 
banks allowed marketing third-party cookies on their websites, and yet only fifteen percent of them 
disclosed this in their privacy policies. We also found that the naming of cookie controls and the labeling of different cookie types were both highly varied, which may confuse consumers and prevent them from effectively opting out.


\section{Discussion}

\subsection{Key Findings and Implications}
Our findings have multiple implications for privacy transparency in the financial sector.

\textbf{Clarity of GLBA notice eroded by additional policies.}
Compared to Cranor et al. \cite{Cranor2016} a decade ago, we found that fewer banks disclose data sharing and more provide required opt-outs in their GLBA notice. 
Our analysis of a more complete set of privacy notices that reflect various current regulatory requirements and disclosure practices also showed that about 44\% of the largest U.S. banks we examined provided at least one other privacy policy in addition to the legally mandated GLBA notice. 
With these additional privacy policies that cover a wider range of data practices related to online and mobile services, the length and complexity of a bank's privacy disclosures increase with bank size.
This proliferation of privacy statements, particularly the coexistence of GLBA notices and other privacy policies, reflects banks’ efforts to comply with an increasingly fragmented regulatory privacy environment. While the GLBA short-form notice was designed to enhance transparency and facilitate compliance, its narrow scope on financial personal information has not kept pace with the digital data practices banks now engage in.
As a result, the GLBA short-form notice, though designed to be concise and user-friendly, may no longer serve as the primary or most informative privacy disclosure for consumers. \looseness=-1

\textbf{Inconsistencies due to GLBA's narrow scope.}
When a bank stated ``no'' for any GLBA sharing purpose, with a 15–27\% likelihood, the same bank made related affirmative sharing statements elsewhere in its other privacy policies, such as allowing third-party marketing or analytics cookies, with marketing/advertising disclosures more common than analytics/research. 
Overall, of 915 banks that provided a GLBA notice and at least one other policy, 53.8\% banks' policies contained at least one such inconsistency. 
Rather than legal noncompliance, this phenomenon is rooted in GLBA's narrow scope on both financial personal information and particular sharing practices.
Yet these legal nuances are likely difficult to recognize for consumers: the GLBA notice uses the same generic term “personal information” as other policies, and offers only five examples to illustrate what the personal information includes without clarifying what falls outside its scope (e.g., online behavioral tracking, location, etc.).
As a result, consumers may reasonably interpret a ``no'' in the GLBA table as a denial of sharing, while other policies reveal otherwise, rendering the GLBA notice uninformative at best and misleading at worst. \looseness=-1

\textbf{Inconsistencies due to CCPA’s protective impact.}
We observed a less concerning but still notable pattern in the opposite direction: when a bank indicated ``yes'' to any GLBA sharing purpose, with a 20-37\% likelihood, the same bank disclosed not sharing with third parties in its other privacy policies.
Many banks explicitly limited such sharing for California residents, which highlights the CCPA’s effectiveness in limiting data exposure for California consumers at least.
Other laws, including GLBA, should also provide stronger protections against third-party sharing beyond requiring transparency.\looseness=-1

\textbf{Limited and unclear privacy opt-outs.}
We found a substantial gap between banks' disclosed third-party sharing and the opt-outs they provided, particularly in relation to CCPA opt-outs and cookie practices. 
Among banks' websites that disclosed allowing third-party marketing or analytics cookies, only 13.3\% implemented cookie controls. These controls also varied widely, with over 40 different cookie labels used. 
Furthermore, many banks' websites used third-party marketing cookies without disclosing it. 
Notably, half of the banks that disclosed the sale or sharing of personal information under CCPA failed to provide the required opt-out link or honor GPC signals. 
These findings demonstrate a lack of regulation and enforcement regarding third-party tracking on banks' websites.

\textbf{Implications for future research.} 
The specific forms of inconsistency we identified across privacy policies under different regulations, together with the different types of sharing statements, may inform future user studies on how overlapping policies and controls affect user understanding as well as the efficacy of policies and controls. Our collection of documents and annotated segments, which we made available as artifacts, could support training classifiers to identify sharing-related disclosures across policy types beyond those covered by existing resources (e.g., OPP-15~\cite{wilson2016creation}). Such tools may additionally support longitudinal monitoring of privacy policy changes, compliance audits of privacy practices, and analyses in other sectors with a more automated approach.

\subsection{Public Policy Recommendations}

Based on our findings, we provide recommendations to enhance the clarity and usability of privacy policies, and consumer protection.

\textbf{Clarify the scope of GLBA notices in template.}
The face-value and potentially misleading inconsistencies between GLBA and other privacy disclosures stem from GLBA’s narrow scope that is not made explicit. The GLBA model notice should be revised to clarify in the “What?” box that the notice \textit{only} applies to financial personal information. If a bank collects and shares data beyond what GLBA covers, this should also be clearly stated.\looseness=-1

\textbf{Expand the data-sharing table.}
Furthermore, the GLBA data-sharing table should be updated to reflect today's data practices, especially online tracking, and broader categories of data sharing. The GLBA model notice could integrate a list of commonly shared personal information types that consumers care about most based on the extensive privacy research conducted since the GLBA notice was developed (e.g., behavioral tracking, location, etc.).

\textbf{Improve machine readability.}
Our automated analysis found frequent formatting issues when processing GLBA PDF notices (e.g., misaligned table columns) due to their visual layout. 
Some GLBA notices were even images, making them unreadable by screen readers. Providing GLBA notices in structured, machine-readable formats would ensure accessibility and facilitate oversight.

\textbf{Prominently reference other privacy policies.}
Some banks used the GLBA notice's ``Other important information'' box to point consumers to a bank's other privacy policies. Yet, it is likely easily overlooked as it appears on the second page at the bottom. Instead, the GLBA model notice could be revised to place structured references to additional disclosures on the \textit{first} page.\looseness=-1

\textbf{Standardize and reconcile privacy policies across laws.}
The multiple privacy policies provided by the \textit{same} bank underscores the urgent need to modernize the layered privacy regulation framework to match current practices. 
Regulators and industry should work toward standardization to reduce redundancy and inconsistency across privacy policies. This includes unifying terminology (e.g., replacing outdated terms like GLBA’s ``non-public information'' with clearer and consistent definitions of ``personal information'')
and harmonizing usable opt-out designs (e.g., modeled after the ``Your Privacy Choices'' CCPA opt-out link).

\textbf{Improve structure and format across policies.}
Unified, rather than policy-specific, disclosure requirements across laws could reduce consumer confusion and lower the cognitive burden of parsing distributed, unstructured, and inconsistent privacy information.\looseness=-1

\textbf{Simplify and centralize opt-out controls.}
Consumers now must navigate multiple interfaces (e.g., website footer, cookie banner, policy text) to opt out of data sharing. Centralizing opt-outs into a single, intuitive interface would reduce friction.
In addition, automated opt-out mechanisms, such as GPC signals, should not only be required to be honored (as is the case with CCPA) but a response from the bank that the signal has been received should also be required (e.g., in HTTP reply or visual indicator on website).


\section{Conclusion}

Our analysis of the 2,073 largest U.S. bank websites revealed a fragmented privacy disclosure environment for consumers.
Nearly half of the banks provided multiple privacy policies, which are long, difficult to read, and especially complex for larger banks.
We identified two key inconsistencies in third-party marketing-related disclosures. The concerning cases are where a bank stated ``no'' to sharing in the GLBA notice but disclosed sharing elsewhere, which may mislead consumers. We also found limited alignment with CCPA-required opt-out controls and inadequate support for third-party cookie opt-outs. These findings highlight the difficulties that the layered and overlapping regulatory requirements bring to both banks and consumers.
To restore transparency, we suggest that privacy disclosure requirements and opt-out controls must be harmonized across laws.\looseness=-1


\begin{acks}
This research has been partially supported by the National Science Foundation under Award No. 2105734; Van Tran has additionally been supported by the National Science Foundation under Award No. 2334996. 
We thank Shomir Wilson, Shahriar Shayesteh, Kirsten Martin, Mukund Srinath, Maaz Bin Musa, Aysun Öğüt, Eera Bhatt, Lee Matheson, and Sumit Asthana for their input and feedback. 
The authors used Grammarly for proofreading.
\end{acks}

\bibliographystyle{ACM-Reference-Format}
\balance
\bibliography{bibli-local}



\end{document}